\documentclass[prb,twocolumn,superscriptaddress]{revtex4-1}

\usepackage{amsmath,amsfonts}
\usepackage{graphicx}
\usepackage{enumerate}

\newcommand{\e}{\mathrm{e}}
\newcommand{\up}{\uparrow}
\newcommand{\dw}{\downarrow}
\newcommand{\ket}[1]{\left| #1 \right\rangle}

\newcommand{\braket}[2]{\left\langle #1 \right. \! \left| #2 \right\rangle}

\allowdisplaybreaks

\date{\today}

\begin{document}



\title{Magnetic-field switchable metal-insulator transitions in a quasi-helical conductor}

\author{Bernd Braunecker}
\affiliation{Departamento de F\'{\i}sica Te\'{o}rica de la Materia Condensada,
Facultad de Ciencias, Universidad Aut\'{o}noma de Madrid, 28049 Madrid, Spain}
\author{Anders Str\"{o}m}
\affiliation{Department of Physics, University of Gothenburg, SE 412 96 Gothenburg, Sweden}
\author{G. I. Japaridze}
\affiliation{Andronikashvili Institute of Physics, Tamarashvili 6, 0177 Tbilisi, Georgia}
\affiliation{Ilia State University, Cholokashvili Avenue 3-5, 0162 Tbilisi, Georgia}

\begin{abstract}
We study Anderson localization in disordered helical conductors that are obtained
from one-dimensional conductors with spin-orbit interaction and a magnetic field,
or from equivalent systems. We call such conductors ``quasi-helical'' because
the spins of the counterpropagating
modes are not perfectly antiparallel and have a small spin-wavefunction overlap
that is tunable by the magnetic field.
Due to the overlap, disorder backscattering is possible and allows a
localization transition.
A conductor can pass through two localization transitions with increasing field,
one from the conventionally localized system to
the quasi-helical conductor (with localization length exceeding the system length),
and one at a higher field again to a localized state, due
now, however, to backscattering below the magnetic field induced pseudo-gap.
We investigate these transitions using a unified two-step renormalization group approach.
\end{abstract}


\maketitle

\section{Introduction}

Over 50 years ago, Anderson showed that a metal-insulator transition
can arise due to localization of particles by scattering on a disorder
potential.\cite{anderson:1958} Since then, Anderson localization
has evolved into an important
topic of condensed matter physics, photonics, and ultracold atom
gases.\cite{localization} While the basic localization mechanism
can be understood on a single-particle basis, a complete investigation
needs to include further interactions,
especially if they compete against localization.
To study such a competition it would be advantageous
to control the interactions externally, and to pass through the metal-insulator
transition on demand.
Here we show that this can indeed be achieved
in the presently much investigated one-dimensional (1D) helical conductors
through an external magnetic field.

Helical conductors are characterized by spin-filtered transport,
in which opposite spins (or Kramers partners) are bound to the
right ($R$) and left ($L$) moving conduction modes.
In conventional 1D conductors, even weak disorder is already sufficient
to turn the conductor into an insulator\cite{Giamarchi,GiamarchiSchulz}
by disorder-induced backscattering between the Fermi points $\pm k_F$.
In a helical conductor, however, backscattering is only possible together with
a spin-flip, and the conductor is insensitive to normal, spin-preserving disorder
scattering.\cite{disorder_TI}

Helical conductors appear at the edges of
topological insulators,\cite{TI} or in quantum wires\cite{streda:2003} or nanotubes\cite{klinovaja:2011} 
with strong spin-orbit interaction (SOI).
They have attracted much attention recently as they allow for
spin-filtering, \cite{streda:2003} Cooper pair splitting, \cite{sato:2010:2011}
and, if in contact with a superconductor, the realization of
Majorana end states.\cite{majorana}
While localization cannot occur in a perfect helical conductor,
many of the investigated conductors are imperfect and we call them
``quasi-helical'':
the spins moving in opposite directions are not perfectly antiparallel.
They provide the handle to tune localization externally.
Examples are semiconductor nanowires with strong SOI in the presence of
a uniform magnetic field\cite{streda:2003} or, without SOI, of a spiral
magnetic field,\cite{braunecker:2010} which can also take the form of
a spiral Overhauser field due to ordered nuclear spins.\cite{braunecker:2009}

\begin{figure}
	\begin{center}
		\includegraphics[width=\columnwidth]{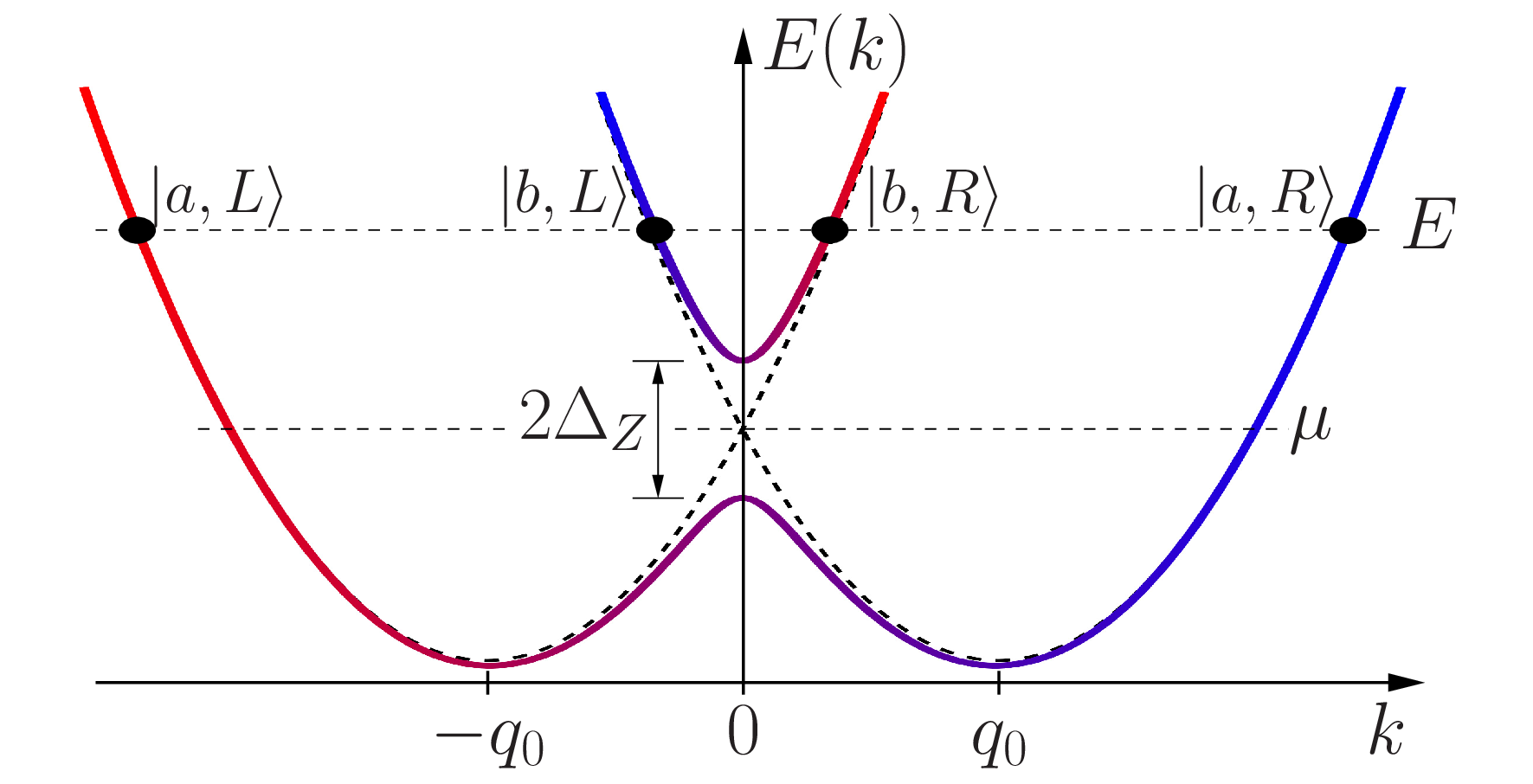}
		\caption{Electron dispersion for the noninteracting quantum wire with SOI
		for $B_x = 0$ (dashed lines) and $B_x \neq 0$ (solid lines). The magnetic field opens a
		pseudo-gap $\Delta_Z = \mu_B g |B_x|/2$ at $k=0$.
		We denote the resulting lower (upper) band by $a$ ($b$). At any energy $E$ the left ($L$)
		and right ($R$) moving states are labeled by $\ket{a/b, L/R}$ as indicated in the figure.
		The color gradient from red to blue indicates the spin mixing by $B_x$.
		}
		\label{bands}
	\end{center}
\end{figure}

The SOI shifts the spin $\up,\dw$ bands by the wave vectors $\pm q_0$
(Fig.~\ref{bands}, dashed lines).
A magnetic field $B_x$ perpendicular to the $\up,\dw$ axis has now two effects.
It lifts the degeneracy at $k=0$ by opening a pseudo-gap (Fig.~\ref{bands}, solid lines),
and it breaks time-reversal symmetry. Tuning the Fermi level to the center of the pseudo-gap
by letting $k_F=q_0$ allows conduction only through the modes close to momenta $\pm 2q_0$
with opposite spins. Through $B_x$, however, these spins are no longer antiparallel,
disorder backscattering becomes again possible, and localization can occur.

In this paper we provide a unified approach to such localization, taking into
account disorder, magnetic field, spin overlaps, and electron interactions.
We formulate a two-step renormalization group (RG) approach within the
Luttinger liquid (LL) framework that provides us with a transparent picture of the
underlying physics.
We consider the zero temperature case, valid if the temperature is smaller than any of the gap sizes.

As a result, we find that varying $B_x$ can cause localization transitions of two
kinds. A localized system\cite{Giamarchi,GiamarchiSchulz} at $B_x=0$
can make a transition to a quasi-helical conductor at a critical field
$B_x^*$ lying (for the example of InAs nanowires) in the range
up to $\sim 1$ T. This transition is quite abrupt and appears when the $B_x$ generated pseudo-gap
overcomes the disorder gap, and the quasi-helical state is lower in energy. 
The transition is marked by a rapid increase of the localization length $\xi_\text{loc}$,
and the system is conducting if $\xi_\text{loc}$ is greater than the system length $\mathcal{L}$. 
Through the equal-spin overlap, $\xi_{\text{loc}}$ then becomes strongly $B_x$ dependent 
and decreases with increasing field. At a larger critical field $B_x^{**} > 1$ T, 
when $\xi_\text{loc} < \mathcal{L}$, 
backscattering between $\pm 2 q_0$ causes again localization,
resulting in a $4q_0$ modulated density wave coexisting
with the transverse spin polarization generated by $B_x$.
This coexistence of density wave and uniform polarization distinguishes
this phase from conventional localization, and we shall call it ``sub-gap localization.''
For very strong disorder, the helical phase is suppressed ($B_x^*=B_x^{**}$),
and a transition takes place directly between the two localized phases.
We show concrete results for InAs nanowires in Sec. \ref{sec:results} below. 
The underlying physics, however,
is general and applies to many materials, except for the edge states of topological
insulators due to their different band structure. For the latter, quasi-helicity
can still be obtained, but has different interesting consequences.\cite{soori:2011}

The plan of the remainder of the paper is as follows.
In Sec. \ref{sec:model} we introduce the model of the quasi-helical conductor.
The approach for its solution in the presence of disorder is discussed in 
Sec. \ref{sec:approach}. In Sec. \ref{sec:bosonization} we present the 
necessary background of the bosonization framework, and in Sec. \ref{sec:RG} we
discuss the technical details of the two-step RG approach. Section \ref{sec:results}
contains the discussion of the results for the example of InAs nanowires and 
the conclusions. 

\begin{figure}
	\begin{center}
		\includegraphics[width=\columnwidth]{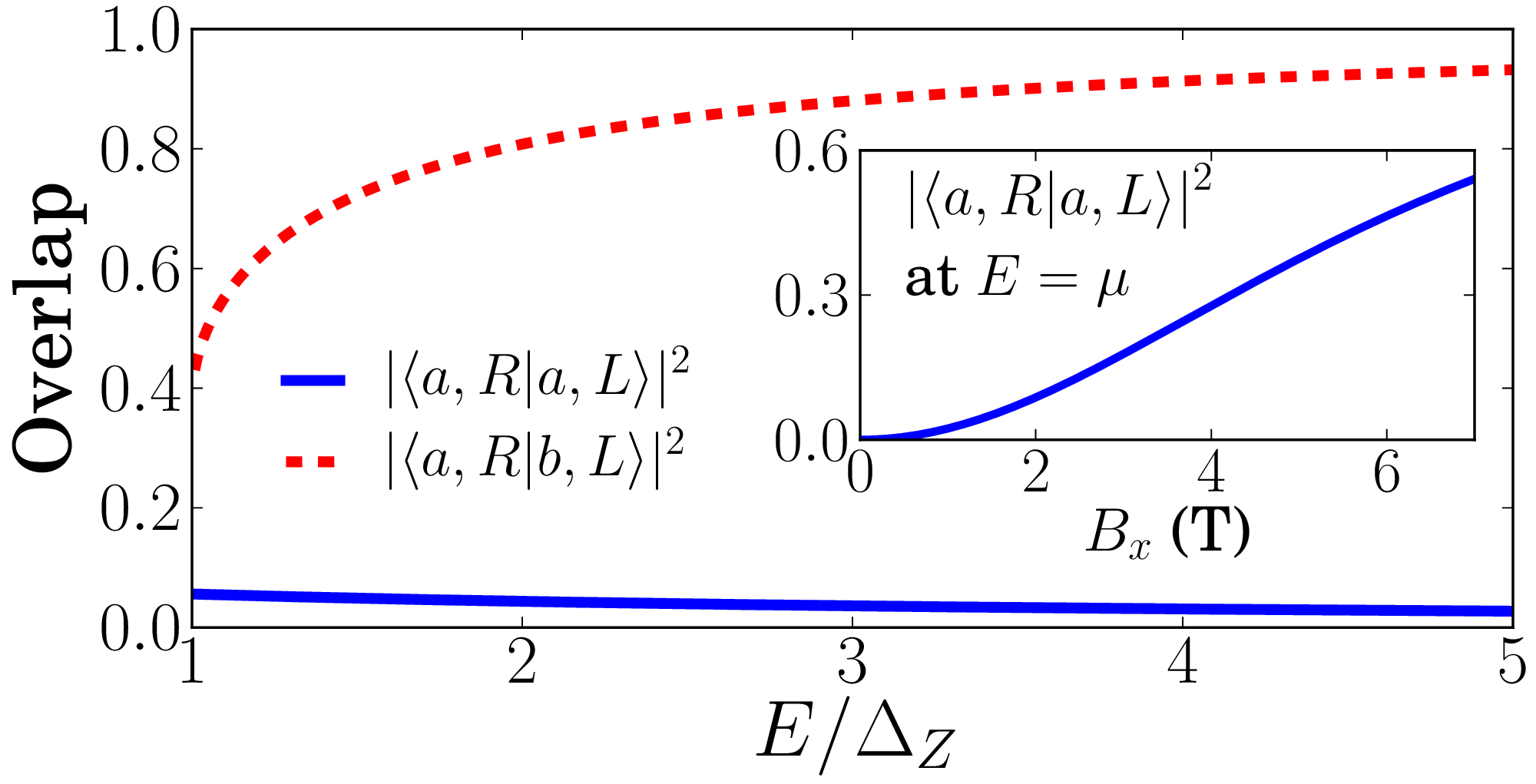}
		\caption{Overlap integrals $|\braket{a,R}{a,L}|^2$ and $|\braket{a,R}{b,L}|^2$
		as functions of energy $E>\Delta_Z = \mu_B g |B_x|/2 \approx 0.5$ meV (for InAs with
		$B_x = 2$ T).\cite{braunecker:2010}
		The states $\ket{a/b, L/R}$ are as indicated
		in Fig. \ref{bands}.
		The inset shows the $B_x$ dependence of $|\braket{a,R}{a,L}|^2$ at energy $E=\mu$
		in the middle of the $\Delta_Z$ gap.
		\label{overlap}}
	\end{center}
\end{figure}


\section{Model}
\label{sec:model}

We consider a generic interacting 1D quantum wire
of length $\mathcal{L}$ with SOI and an external magnetic field,
described by the Hamiltonian
\begin{align}
	H = &\sum_{s,s'} \int dx \, \psi_{s}^\dagger(x)
	\biggl[
		\left(\frac{p^2}{2 m}-\mu\right) \delta_{s,s'}
		+
		\alpha_R\sigma^z_{s,s'} p
\nonumber \\
		&+
		\Delta_Z \sigma^x_{s,s'}
	\biggr]
	\psi_{s'}(x)
	+ U_{e-e}
	+ U_{dis}.
\label{eq:H}
\end{align}
Here $\psi_s(x)$ are the electron operators at position $x$ for spin $s = {\up,\dw} = +,-$,
$p = -i\hbar \partial_x$ is the momentum operator, $\mu$ the chemical
potential, $m$ the band mass,
$\alpha_R$ the Rashba SOI strength, and
$\Delta_Z = \mu_B g B_x/2$ the Zeeman interaction strength,
with Bohr magneton $\mu_B$ and Land\'{e} $g$-factor $g$.
$U_{e-e}$ is a general electron-electron interaction and
$U_{dis}$ the disorder potential.
$\sigma^{x,z}$ are the spin Pauli matrices, with the spin axes chosen such that
$\alpha_R$ couples to $\sigma^z$ and $B_x$ to $\sigma^x$.
Equation \eqref{eq:H} describes a quantum wire with a single transverse
subband, and we exclude the influence of higher subbands.\cite{governale,schulz:2009}
Realizations of Eq. \eqref{eq:H} are found in GaAs, InAs,\cite{braunecker:2010,nadj-perge:2010,shtrikman:2011}
Ge/Si,\cite{kloeffel:2011,hao:2010} or InSb nanowires.\cite{nadj-perge:2012}

Without $U_{e-e}$ and $U_{dis}$, Eq. \eqref{eq:H} leads to the bands shown in Fig. \ref{bands}.
The SOI shifts the $\up,\dw$ dispersions by $\pm q_0 = \pm m \alpha_R/\hbar$.
The $B_x$ field opens a pseudo-gap at momentum $k=0$ by spin-flip scattering.
We denote the lower (upper) resulting band by $a$ ($b$).
Diagonalizing $H$ leads to the dispersions
$E_k^{a,b} = \hbar^2k^2/2m \mp \sqrt{\Delta_Z^2+\hbar^2\alpha_R^2 k^2}$, and the
wave functions
$\ket{a,k} = u_{-k} \ket{\up} - u_k \ket{\dw}$,
$\ket{b,k} = u_k    \ket{\up} + u_{-k} \ket{\dw}$,
with
$u_k = \left[1+\hbar \alpha_R k/\sqrt{\Delta_Z^2+\hbar^2\alpha_R^2 k^2}\right]^{1/2}/\sqrt{2}$.
Overlaps of the form $\braket{a/b,k}{a/b,k'}$ weight the disorder backscattering
amplitudes at a given energy. Figure \ref{overlap} shows the two overlaps
which will be relevant for the localization transition.


\section{Approach}
\label{sec:approach}

To study a disordered interacting conductor in 1D, it is generally
convenient to bosonize Eq.~\eqref{eq:H} and use an RG approach.
For the present model, however, this approach exhibits some peculiarities.
First, the overlap integrals depend on $k$ and the problem is no longer scale
invariant, a central assumption for RG. Second, the low-energy physics below the
scale set by $\Delta_Z$
is described mostly by the band $a$. Hence, the RG procedure of reducing the energy scale
can lead to a problematic crossing the bottom of the $b$ band, which also
invalidates the bosonization formulation that neglects
the band curvature.
While the exact treatment of every aspect would require a
separate investigation, the following two observations allow us to still implement a global RG scheme
capturing the relevant physics in a transparent way.

First, Fig.~\ref{overlap} shows that the spin overlaps above $\Delta_Z$ approach their asymptotics
only slowly with increasing energy $E$
(as $|\braket{a,R}{a,L}|^2 \sim \Delta_Z^2/2 m\alpha_R^2 E$
and
$|\braket{a,R}{b,L}|^2\sim 1-\Delta_Z^2/4 E^2$).
Considering constant overlaps evaluated, e.g., at half of the high energy cutoff scale
yields a valid, rather conservative estimate of the true influence of the overlaps.
Second, the interactions also renormalize $\Delta_Z$ and it grows quickly under
the RG flow.\cite{braunecker:2009,braunecker:2010}
Hence the reduced RG bandwidth $E$
meets the growing $\Delta_Z$ at a value $E=\Delta_Z^*\gg \Delta_Z$ well above
the bottom of the $b$ band, at which the linearity of the
bare dispersion remains valid.

Due to these two properties, we propose a unified two-step RG approach to the effect of disorder
based on the LL theory. In the first step, we integrate over
high energies far above $\Delta_Z$ using a constant backscattering overlap
and the standard inclusion of the disorder potential.\cite{Giamarchi,GiamarchiSchulz}
If the disorder is strong enough,
localization already occurs in this regime by the conventional
backscattering mechanism.
Otherwise, at $E=\Delta_Z^*$, we proceed to the second step.

Crossing the gap $\Delta_Z^*$ corresponds to freezing out the interaction-generated
density fluctuations that renormalize $\Delta_Z$, and hence is not a singular
transition through the true band bottom. While the description of the proper
transition may be quite challenging, the physics below $\Delta_Z^*$ becomes again simple.
It is described by a different LL theory
for the modes originating from the $a$ band only.\cite{braunecker:2009,braunecker:2010,braunecker:2012}
In this regime it is legitimate to use a larger spin overlap value and, within the accuracy of LL
and RG theories, we choose to take the overlaps at the chemical potential $\mu$
(see Fig. \ref{overlap}, inset).
The evaluation of the disorder backscattering (the second RG step) follows then the standard lines
of a spinless LL, and localization occurs if $\xi_\text{loc} < \mathcal{L}$.


\section{Bosonization}
\label{sec:bosonization}

Above the pseudo-gap, all fluctuations of the $a$ and $b$ bands
must be taken into account.
As we account for the overlap integrals separately, the standard
$s=\up,\dw$ basis rather than the $a,b$ basis\cite{gangadharaiah:2008}
is more convenient for bosonizing.
The electron operators are decomposed into $R$ and $L$ moving components
as $\psi_{s}(x) = \e^{i k_{FRs} x} \psi_{R,s}(x) + \e^{-i k_{FLs} x} \psi_{L,s}(x)$,
for $k_{FRs},k_{FLs}$ the two Fermi points of the (nominal) spin $s$ bands.
We then write $\psi_{r,s}(x) = \eta_{r,s} \exp\left(r i \sqrt{\pi}[\varphi_s(x)+ r \vartheta_s(x)]\right)/\sqrt{2\pi \kappa}$
for $r = R,L=+,-$, the boson fields $\varphi_s$ and $\vartheta_s$ satisfying
$[\varphi_{s'}(x'),\partial_x\vartheta_{s}(x)] = i \delta_{s,s'} \delta(x-x')$,
the Klein factors $\eta_{r,s}$, and the short distance cutoff
$\kappa$.
Defining  $\varphi_{\rho},\vartheta_{\sigma} = (\varphi_\uparrow\pm \varphi_\downarrow)/\sqrt{2}$ and
$\vartheta_{\rho},\varphi_{\sigma} = (\vartheta_\uparrow\pm \vartheta_\downarrow)/\sqrt{2}$, we obtain the Hamiltonian\cite{gangadharaiah:2008}
(for $U_{dis}=0$)
$H_{bos} = \sum_{\nu=\rho,\sigma} (u_\nu/2) \int dx \left[ K_\nu^{-1} (\partial_x \varphi_\nu)^2 + K_\nu (\partial_x \vartheta_\nu)^2 \right]$,
where $K_{\rho,\sigma}$ measure the interaction strengths ($0<K_\rho<1$ for repulsive $U_{e-e}$ and\cite{gangadharaiah:2008}
$K_\sigma \approx 1$) and $u_{\rho,\sigma}$ are renormalized velocities.
Because $\mu$ lies in the pseudo-gap, $\Delta_Z$ renormalizes as well,
expressed by a relevant Hamiltonian $\propto \Delta_Z \cos(\sqrt{2\pi}(\phi_\rho+\theta_\sigma))$,
acting on the fields of the $b$ band.\cite{braunecker:2009,braunecker:2010}
We neglect further existing Cooper scattering processes,\cite{gangadharaiah:2008} which are
overruled by the renormalization of $\Delta_Z$ and the disorder here but otherwise would
dominate the physics.

The second RG step starts
when the growing $\Delta_Z$ meets the reduced bandwidth $E$
at $E = \Delta_Z^*$. The $b$-band fields are then fully gapped,\cite{braunecker:2009,braunecker:2010}
and the low-energy theory is described by the fields related to the $a$ band alone.
The corresponding Hamiltonian is obtained from $H_{bos}$ by suppressing all $b$-related fields,
which leads to\cite{braunecker:2009,braunecker:2012}
$H_{bos}^a = (u_a/2) \int dx \left[ K_a^{-1} (\partial_x \varphi_a)^2 + K_a (\partial_x \vartheta_a)^2 \right]$,
with $u_a^2 = (1/4) \left[u_\rho^2+u_\sigma^2 + u_\rho u_\sigma (K_\rho K_\sigma+ K_\rho^{-1}K_\sigma^{-1})\right]$
and $K_a^2 = K_\rho K_\sigma^{-1} (u_\rho K_\rho+u_\sigma K_\sigma^{-1})/(u_\rho K_\sigma^{-1}+u_\sigma K_\rho)$.
We have neglected here also a marginal coupling between $a$ and $b$ fields\cite{braunecker:2009}
because it affects mostly the fermionic response\cite{braunecker:2012,schuricht:2012} but not
the bosonic theory.


\section{Disorder and Renormalization Group approach}
\label{sec:RG}

Disorder can be expressed by a random potential $U_{dis}$
with Gaussian distribution\cite{Giamarchi,GiamarchiSchulz} that scatters between the
bands $i,j=(a/b,L/R)$.
The scattering amplitude is described by the dimensionless disorder strength $\tilde D_{ij}$, proportional
to the square of the strength of each individual scattering potential, which in turn is proportional to
$\left|\braket{i}{j}\right|^2$, hence $\tilde{D}_{ij} = \left|\braket{i}{j}\right|^4 \tilde{D}$.
Following the standard replica disorder averaging approach,\cite{Giamarchi,GiamarchiSchulz} we obtain the scaling equations
(including also the amplitude $y$ expressing bulk backscattering, see Ref. \onlinecite{Giamarchi})
\begin{align}
	\partial_l K_\rho			&= -u_\rho K_\rho^2(2 \tilde D_{ab}+\tilde D_{aa})/4u_\sigma,\\
	\partial_l K_\sigma		&= -K_\sigma^2 (\tilde D_{ab} + y^2 )/2+\tilde D_{aa}/4,\\
	\partial_l y			    &= (2-2K_\sigma)y-\tilde D_{ab},\\
	\partial_l \tilde D_{aa}	&= (3-K_\rho-K_\sigma^{-1})\tilde D_{aa},\\
	\partial_l \tilde D_{ab}	&= (3-K_\rho-K_\sigma-y)\tilde D_{ab},\\
	\partial_l u_\rho			&= -K_\rho u_\rho^2(2\tilde D_{ab}+\tilde D_{aa})/4u_\sigma,\\
	\partial_l u_\sigma		    &= - u_\sigma K_\sigma \tilde D_{ab}/2- u_\sigma K_\sigma^{-3}\tilde D_{aa}/4,\\
	\partial_l \delta(l)        &= \left[2-(K_\rho+K_\sigma^{-1})/2\right] \delta,
\end{align}
with $l$ the running RG scale,
$\tilde{D}_{aa}=\tilde{D}_{(a,R),(a,L)}$, and $\tilde{D}_{ab}=\tilde{D}_{(a,R),(b,L)}$.
We have neglected scattering between $(b,R) \leftrightarrow (b,L)$,
$(a,R) \leftrightarrow (b,R)$, $(a,L) \leftrightarrow (b,L)$ after verifying that
it has no effect.
We have also defined $\delta(l) = \Delta_Z(l)/ E(l)$
with $E(l) = \hbar v_F/ \kappa(l)$ the running effective bandwidth, for $\kappa(l) = \kappa \e^{l}$
and Fermi velocity $v_F$.\cite{braunecker:2009}
The latter equations express the competition between disorder backscattering and the delocalizing
effect by repulsive interactions\cite{Giamarchi} and by $B_x$ induced spin-flip scattering.
Localization occurs when the latter is not strong enough
such that $\tilde{D}_{ab}\sim 1$  ($\tilde{D}_{aa}$ remains small above the gap)
before we reach $E=\Delta_Z^*$ ($\delta(l) \sim 1$) or
$\kappa(l) > \mathcal{L}$. Otherwise we switch to the second step,
described by $H_{bos}^a$, with the parameters $K_a$ and $u_a$ obtained from the
resulting $K_{\rho,\sigma}, u_{\rho,\sigma}$ of the first step.
As argued above, the spin overlap weighting $\tilde{D}_{aa}$
can now (discontinuously) be replaced by $\braket{a,L}{a,R}$ evaluated at $E=\mu$.
The RG equations are
\begin{align}
	\partial_l K_a &= - K_a^2 \tilde D_{aa}/2,\\
	\partial_l \tilde{D}_{aa} &=  (3-2K_a) \tilde D_{aa},\\
	\partial_l u_a &= - u_a K_a\tilde D_{aa}/2,
\end{align}
describing, for effectively spinless fermions, the competition
between localization and delocalizing superconducting fluctuations.\cite{Giamarchi}
The latter overrule disorder localization above the critical attractive interaction
strength\cite{Giamarchi} $K_a > 3/2$.
However, $K_a<1$ for repulsive interactions,
disorder scattering is relevant, and localization occurs
if $\tilde{D}_{aa} \sim 1$ is reached while
$\kappa(l)<\mathcal{L}$; otherwise the (finite)
system is a helical conductor.


\begin{figure}
	\begin{center}
		\includegraphics[width=1.0\columnwidth]{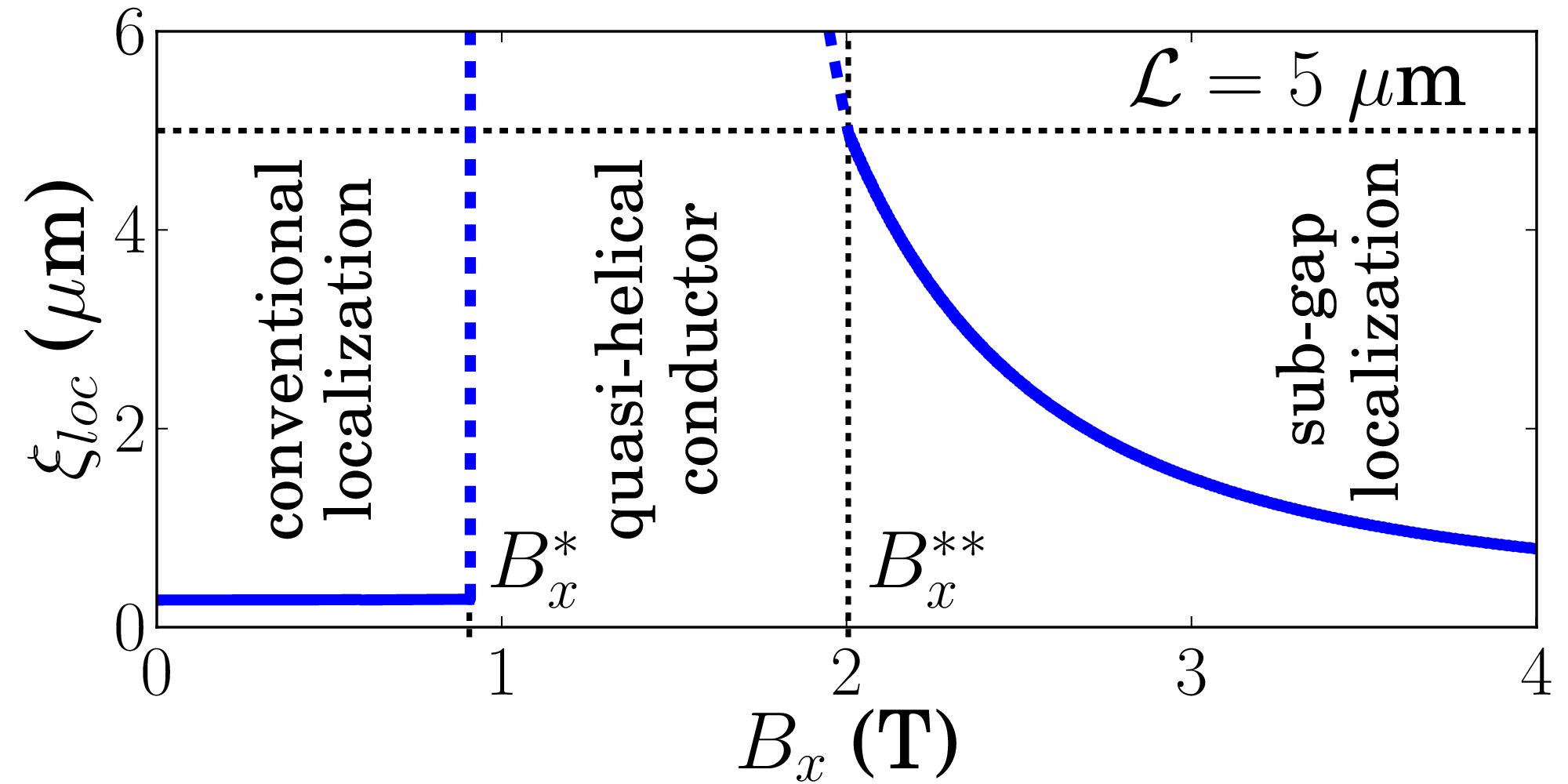}
		\caption{Localization length $\xi_\text{loc}$ of an InAs nanowire as a function
		of $B_x$ for $\tilde{D}=0.01$ and $\mathcal{L} = 5$ $\mu$m.
		Close to $B_x=0$ the system is localized due to conventional backscattering.
		At $B_x \sim B_x^* \approx 0.9$ T the system crosses over to a quasi-helical conductor
		and at $B_x \sim B_x^{**} \approx 2$ T to a localized phase due to sub-gap backscattering.
		The transition at $B_x^*$ results from the competition of the renormalization of $\Delta_Z$ and
		$\tilde{D}_{ab}$ and is thus expected to be quite abrupt
		and independent of $\mathcal{L}$.
		The transition at $B_x^{**}$ is a strongly $\mathcal{L}$ dependent cross-over.
		\label{loclength}}
	\end{center}
\end{figure}
\begin{figure}
	\begin{center}
		\includegraphics[width=1.0\columnwidth]{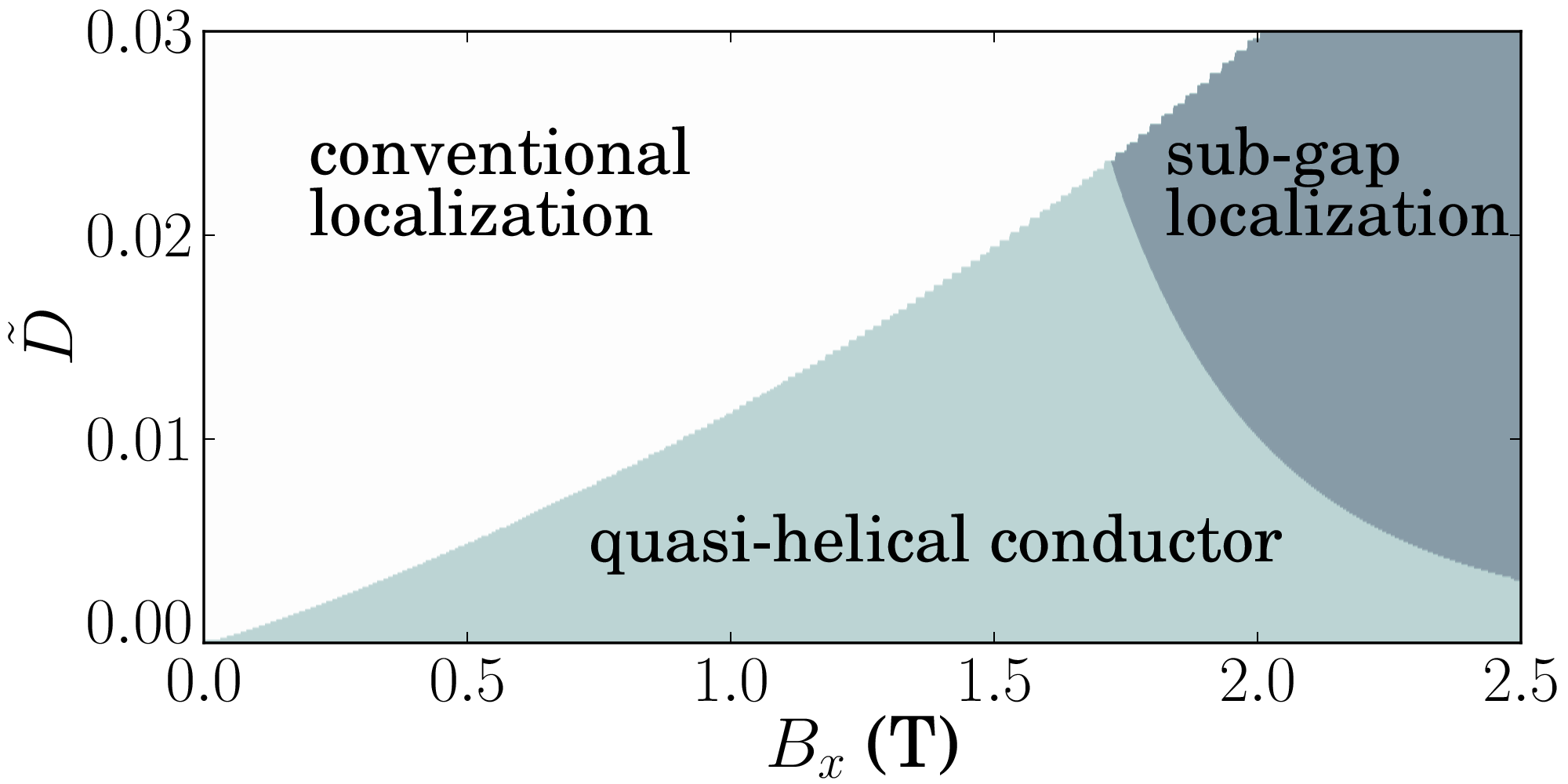}
		\caption{Phase diagram for InAs nanowires
		obtained from curves as in Fig. \ref{loclength}
		for various $\tilde{D}$. At $\tilde{D} > 0.023$
		the quasi-helical phase is suppressed and there is a direct crossover between
		the localized phases. Not visible on the shown scale is the localization threshold
		for $B_x=0$, occurring at $\tilde{D} \approx 0.003$.}
		\label{phase_diag}
	\end{center}
\end{figure}


\section{Results and discussion}
\label{sec:results}

To give a definite example, we focus on InAs nanowires,\cite{Fasth,braunecker:2010} using
$\alpha_R = 4 \times 10^{-11}$ eV m, $g = -9$, $m = 0.040 m_e$ (with electron mass $m_e$),
$v_F = 2 \times 10^{5}$ m/s, $u_\rho=v_F/K_\rho$ and $u_\sigma=v_F/K_\sigma$,
with $K_\rho = 0.5$, $K_\sigma = 1$, $y=0.1\left|\braket{aL}{bR}\right|^2$,
and short length cutoff $\kappa=15$ nm (which is longer than the lattice constant $a_0 =6.06$ \AA,
and expresses a smaller effective bandwidth).
The localization length is $\xi_\text{loc} = \min_{ij}\kappa(l^*)/\tilde{D}_{ij}(l^*)$
with $l=l^*$ the scale at which the RG flow stops (in the first or the second step).
If a $\tilde{D}_{ij}(l^*) = 1$ is reached before $\kappa(l)>\mathcal{L}$, the
system is localized, $\xi_\text{loc} < \mathcal{L}$.
We consider disorder strengths about $\tilde{D} \sim 0.01$, leading at $B_x=0$ to
$\xi_\text{loc} \sim 0.3$ $\mu$m. For a sample length of, e.g., $\mathcal{L} = 5$ $\mu$m,
the system is well localized.
At small fields, the $\tilde{D}_{ab}$ and $B_x$ scattering processes compete,
and if $B_x$ passes a critical value $B_x^* < 1$ T,
$\Delta_Z(l)$ overrules the disorder backscattering and the system becomes a quasi-helical conductor,
where $\xi_\text{loc}>\mathcal{L}$ is now determined by the sub-gap disorder
strength $\tilde{D}_{aa}$.
At $B_x=B_x^{**} \sim 2$ T, $|\braket{a,R}{a,L}|$ becomes large
enough such that $\xi_{loc} < \mathcal{L}$ and the system crosses over into the sub-gap localized phase.
In Fig. \ref{loclength} we plot this crossover behavior for $\tilde{D} = 0.01$. Tracing similar curves for
various disorder strengths $\tilde{D}$ leads to the phase diagram shown in Fig. \ref{phase_diag}.
For strong disorder, the quasi-helical conduction phase is absent, and the
crossover takes place directly between the two localized phases.

To conclude, we note that the disordered quasi-helical system shows at $B_x^*$ a
phase transition from the conventional Anderson localized phase in which $\xi_\text{loc}$
only weakly increases with $B_x$, to a phase in which $\xi_\text{loc}$ jumps to a very 
large value and then decreases with increasing $B_x$. For fields $B_x^* < B_x < B_x^{**}$, 
$\xi_\text{loc}$ exceeds $\mathcal{L}$ and forms a quasi-helical
conductor before crossing over into another localized phase. While the former localized phase
is characterized by the conventional $2k_F =2q_0$ modulated charge density wave, 
the latter sub-gap localized phase combines a $4q_0$ density wave with a uniform electron 
polarization as the order parameter. The existence of a quasi-helical conduction state between
the two insulating phases as a function of $B_x$ is a peculiar behavior that could be used
to test if a conductor is helical. 
Remarkably the conducting phase extends into the regime of quite strong disorder,
which indicates that a quasi-helical conductor does not necessarily require ultraclean samples.
The boundaries of this phase in the phase diagram are controlled predominantly by $\Delta_Z$, and so the
$g$-factor of the material has the largest influence on the phase diagram. For instance,
a similar phase diagram as Fig. \ref{phase_diag} for InSb with\cite{nadj-perge:2012} $g \approx 50$
spans over $B_x = 0 - 0.1$ T and $\tilde{D} = 0 - 0.002$.


\begin{acknowledgments}
We thank F. Assaad, D. Baeriswyl, H. Johannesson, and P. Recher for valuable discussions.
B.B. acknowledges the support by the EU-FP7 project SE2ND [271554]. A.S. acknowledges the
support by the Swedish research council, Grant No.~621-2011-3942. G.I.J. acknowledges the
support by the Georgian NSF Grant No.~ST09/4-447 and by the SCOPES Grant IZ73Z0-128058.
\end{acknowledgments}


\end{document}